# Transformations of Phosphorus under Pressure from Simple Cubic to Simple Hexagonal Structures via Incommensurately Modulations: Electronic Origin

Degtyareva Valentina

Institute of Solid State Physics, Russian Academy of Sciences, Chernogolovka 142432, Russia; Correspondence: degtyar@issp.ac.ru

**Abstract:** The lighter group-V element phosphorus forms the As-type ($hR2$) structure under pressure, above 5 GPa, and at 10 GPa transforms to the simple cubic structure ($cP1$), similar to arsenic. Despite of its low packing density, the simple cubic structure is stable in phosphorus over a very wide pressure range up to 103 GPa. On further pressure increase, the simple cubic structure transforms to a simple hexagonal structure ($hP1$) via a complex phase that was solved recently as incommensurately modulated. Structural transformations of phosphorus are connected with the changes of physical properties. Above 5 GPa phosphorus shows superconductivity with $T_c$ reaching ~9.5K at 32GPa. The crystal structures and properties of high-pressure phases for phosphorus are discussed within the model of the Fermi sphere and Brillouin zone interactions.

## 1. Introduction

Recent high-pressure x-ray diffraction studies revealed unusual complex and low symmetry structures in some simple elements [1], including incommensurate modulated (IM) structures. IM structures were found in the elements of group VII (I and Br) and group VI (S, Se and Te). Very recently, an IM structure was found in a light group V element phosphorous in the phase P-IV stable in the pressure range 107–137 GPa between two simplest phases; simple cubic (cP1) and simple hexagonal (hP1) [2–5]. At pressure ~260GPa P-hP1 transforms to the body-centered cubic structure (bcc) that was recently redfound to form superlattice structure cI16 [6]. Structural transformations in phosphorous under pressure are connected with appearance of superconductivity at 5 GPa and increase of $T_c$ to 9.5K at 32 GPa [7].

All IM structures were observed when elements become metallic under pressure. This implies the importance of the two main contributions to the lattice energy: electrostatic (Ewald) and electronic (band structure) energies. The latter can be lowered due to a formation of Brillouin zone planes near the Fermi level and an opening of an energy gap at these planes. This effect is important for stability of so-called Hume-Rothery phases in Cu – Zn and related alloys [8–10]. Under pressure, the band structure energy part becomes more important leading to a formation of complex low-symmetry structures [11].

For a classical Hume-Rothery phase $Cu_5Zn_8$, the Brillouin zone filling by electron states is equal to 93%, and is around this number for many other phases stabilized by the Hume-Rothery mechanism [12–15]. Diffraction patterns of these phases have a group of strong reflections with their scattering vectors $q_{hkl}$ lying near $2k_F$ and the Brillouin zone planes corresponding to these $q_{hkl}$ form a polyhedron that is very close to the Fermi sphere. We consider configurations of the Brillouin zone and Fermi sphere within a nearly-free-electron model in order to analyze the importance of these configurations for the lattice energy. A computer program, BRIZ, has been developed for the visualization of Brillouin zone – Fermi sphere configurations to estimate Brillouin zone filling by electron states [16]. Thus, with the BRIZ program one can obtain a qualitative picture and some quantitative characteristics on how a structure matches the criteria of the Hume-Rothery mechanism.

## 2. Results and Discussion

Experimentally observed transformations in phosphorus on compression go along the sequence including highly symmetrical structures: simple cubic (sc) – simple hexagonal (sh) – body-centered cubic (bcc). This sequence satisfies the requirements to increase under pressure in packing density (0.52 – ~0.60 – 0.68) and in coordination number (6 – 2+6 – 8+6). Energetically this sequence corresponds to increase in Madelung constant 1.760 – 1.775 – 1.7918 [17] leading to decrease in the electrostatic energy. The low-symmetry intermediate phase of phosphorus-IV appears between the simple cubic and the simple hexagonal structures indicating on importance of the band structure energy for stabilization of this unusually complex structure.

### 2.1. Construction of Brillouin-Jones Zone for Phosphorus – IV

The phase P-IV has an incommensurately modulated crystal structure with superspace group $Cmmm$ $(00\gamma)$ $s00$ [4,5], with the modulation wave vector $\gamma = 0.2673$. The basic lattice is base-centered orthorhombic with lattice parameters $a$ = 2.772 Å, $b$ = 3.215 Å, $c$ = 2.063 Å at 125 GPa [4]. Diffraction pattern and suggested structural model are shown in Figure 1. We consider a commensurate approximant with an 11-fold supercell along the c-axis and a modulation wave vector equal to 3/11 = 0.273 which is close to the experimentally observed value of 0.267.

Taking the experimentally found atomic volume 9.19 Å$^3$ and the number of valence electrons for group V element P equal $z$ = 5 we estimate the value of $2k_F$ equal to 5.05 Å$^{-1}$. Within a nearly free-electron model the Fermi sphere radius is defined as $k_F = (3\pi^2 z/V)^{1/3}$, where $z$ is the number of valence electrons per atom and $V$ is the atomic volume. The $2k_F$ position is indicated on Figure 1 (right inset). Atomic shifts due to the modulation result in appearance of satellite reflections and hence in a formation of additional Brillouin zone planes. Reflections have scattering vectors $H = h\mathbf{a}^* + k\mathbf{b}^* + l\mathbf{c}^* + m\gamma\mathbf{c}^* = h\mathbf{a}^* + k\mathbf{b}^* + (l+m\gamma)\mathbf{c}^*$. For the commensurate approximation, $\gamma = 3/11$, an 11-fold supercell has been employed

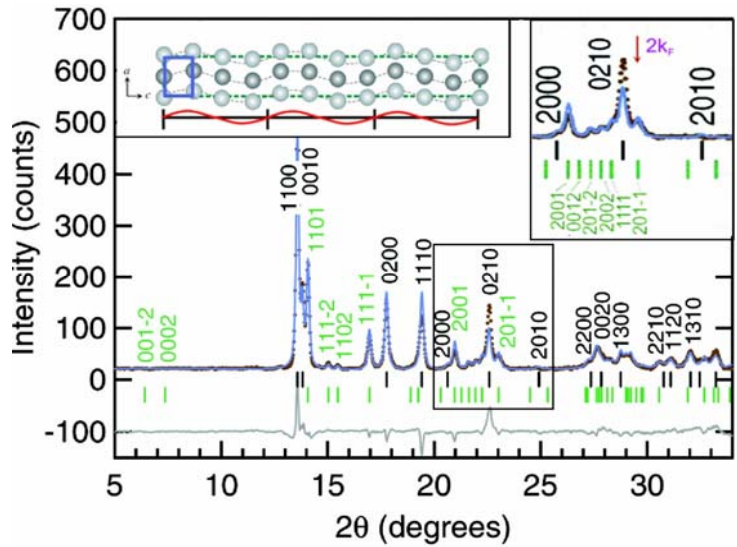

**Figure 1.** The experimental diffraction pattern of P-IV at 125 GPa (from Ref. 4). The tick marks show the peak positions for main (upper) and modulated (lower) reflections. The left inset shows structural model of P-IV in $ac$ plane: the basic cell is shown by solid (blue) line; the commensurate approximant with 11-fold supercell along the c-axis is shown by dashed (green) line. A modulation wave 3/11 is shown below (red). The right inset shows reflections near $2k_F$ indicated by arrow. Indices for reflections are given for the incommensurate cell (4D).

The P-IV basic cell can be considered as an orthorhombic distortion of P-III simple cubic cell. With this distortion some BZ planes are developed a contact with the FS

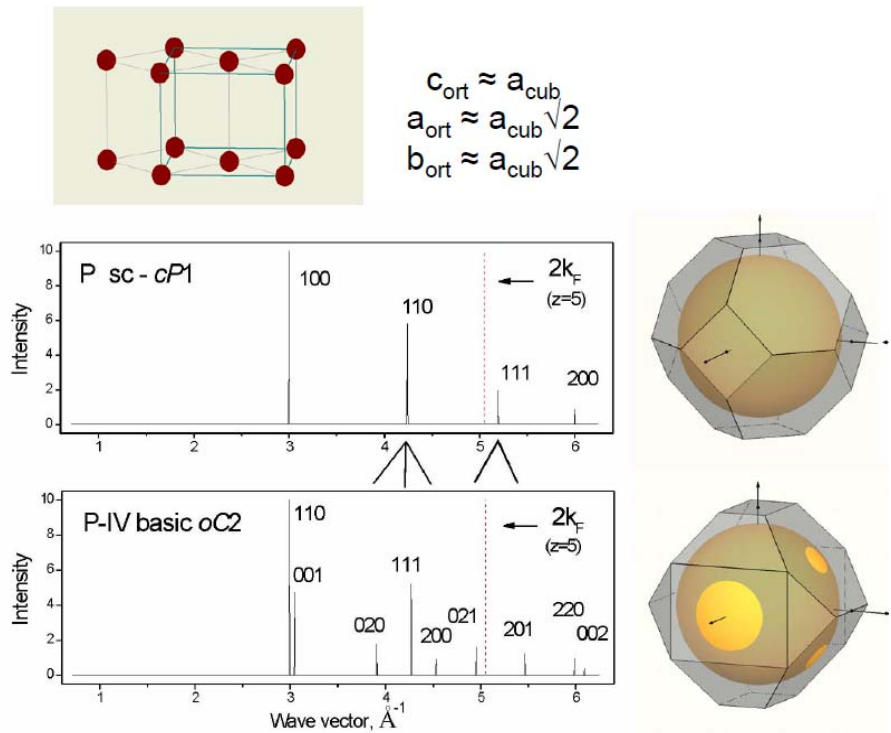

**Figure 2.** Structural relation of P-III *cP*1 and P-IV basic *oC*2 cells (upper panel) Powder diffractions for *cP*1 and *oC*2 structures and FS-BZ configurations for the valence electron number z = 5 are shown in the middle and lower panels.

For modulated structure some additional reflections appeared. We selected a group of reflections near $2k_F$ for construction of the Brillouin-Jones zone (right inset in Figure 1). At first step we constructed planes of reflections (20lm) type and the (0012) reflection as shown in Figure 3. On projection down b* one can see a set of planes lying close to the Fermi sphere. This configuration implies that some deformation of the sphere to an ellipsoid would be necessary. Within the free electron model it was suggested to consider a ratio of the Fermi radius to the ½q (a distance to a plane from the origin) called as "truncation" factor [18] which is usually 1.05 and may increase up to 1.10 as in the case of (2000) plane for phosphorous

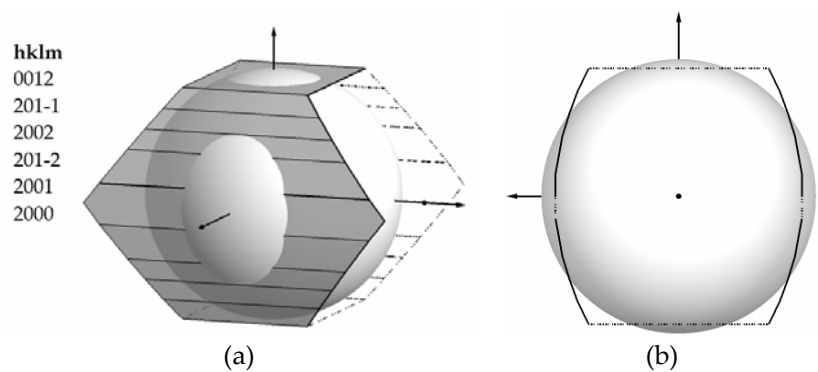

**Figure 3.** Configuration of the Brillouin planes (20*lm*) type and (0012) with the inscribed Fermi sphere for the valence electron number z = 5: (a) common view; (b) the view down b*.

Close to the Fermi sphere are planes (1111) and (0210) that are added to the Brillouin – Jones zone shown in Figure 4. This zone is filled to 86 % by electron states which satisfies the Hume-Rothery mechanism of phase stabilization [16].

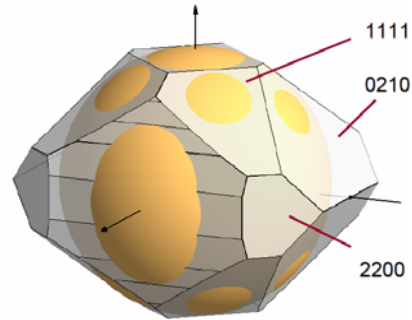

**Figure 4.** The Brillouin – Jones zone for P-IV constructed with planes close to $2k_F$, shown in the inset to Figure 1. Planes (1111), (0210) and (2200) are added to the planes $h0lm$ shown in Figure 3.

Now we consider Fermi sphere – Brillouin zone configuration to find factors defining magnitude of the modulation wave vector. It was noted in Ref. [5] that the $2k_F$ vector is matching the (201-1) peak on the diffraction pattern. In Figure 5 we construct the Brillouin planes (001) and (200) in projection down $b^*$ and mark nodes (001), (200) and (201) of the basic cell. Constructing the $2k_F$ vector through the point where the (200) plane crosses the Fermi sphere we define the position of the (201-1) reflection. The shift of this reflection down the $c^*$ axis from (201) is equal to $\gamma c^*$ and the value of $\gamma$ determined from this construction is 0.268 which is very close to the experimentally found value $\gamma = 0.2673$ [4].

Thus, the plane (201-1) corresponding to the modulation peak (201-1) appears just to cut an empty section where the (200) plane crosses the Fermi sphere. This mechanism is similar to the "nesting" effect and allowed the lowering of electron energy on a Brillouin plane close to the Fermi sphere. Formation of the modulation wave accounts for appearance of other modulation peaks and consequently a group of Brillouin planes near the Fermi sphere as shown in Figures 3 and 4.

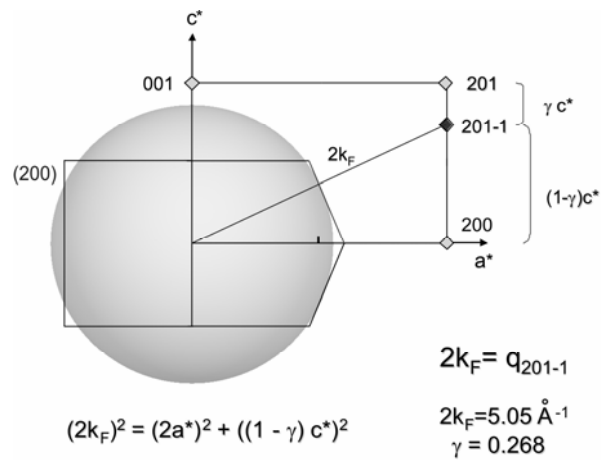

**Figure 5.** Configuration of the reciprocal lattice for P-IV and the Fermi sphere down $b^*$. Three-integer indices correspond to the basic unit cell. The position of the (201-1) reflection is defined by a reciprocal vector equal to $2k_F$ through the point of crossing of the (200) plane with the Fermi sphere. From this construction one can estimate the value of the modulation wave vector $\gamma$.

*2.2. Construction of Brillouin-Jones Zones for Phosphorus –V*

The complex modulated structure of P-IV is intermediate between two the simplest structures: the simple cubic and the simple hexagonal as shown first in [2]. The simple hexagonal phase appears above 137 GPa and has lattice parameters at 151 GPa *a* = 2.1750 and *c* = 2.0628 Å with *c*/*a* = 0.9484.

At this pressure the relative volume is reached 0.444 and the volume decrease at IV-V transition is 5.5% [4] or 4.5% [5]. Theoretical considerations of the electron energy suggested broadening and overlap of electron levels on compression [19]. Thus one may expect for phosphorus as for element of the third period occurrence under pressure a hybridization of 3p and 3d electron levels while the latter is empty in a free atom.

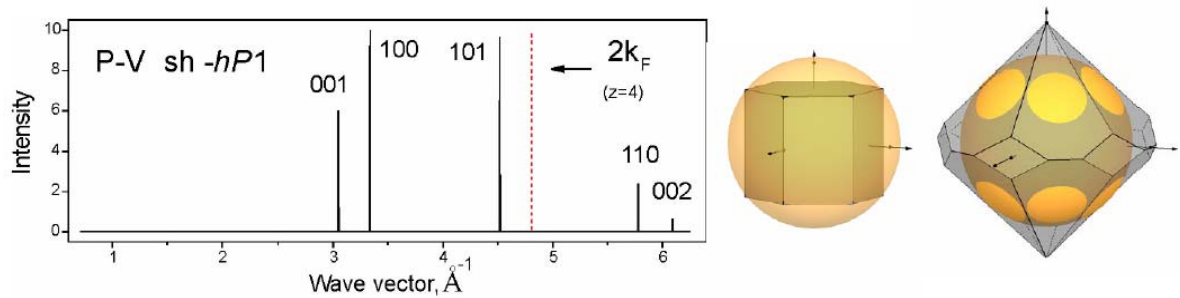

**Figure 6.** The simple hexagonal structure of Phosphorus-V. Diffraction pattern (left) with indication of $2k_F$ position. The Fermi sphere for z = 4 and Brillouin zones of planes: (middle) {100} and {001} types, (right) {101} and {110} types. Data for 151 GPa [2], P-*sh*, *hP*1, SG *P6/mmm a* = 2.175 *c* = 2.0628 Å .

Simple hexagonal structures are known to appear in group-IV elements Si and Ge under pressure [1] and in some binary alloys as In-Sn, Hg-Sn, Cd-Sb, Zn-Sb, Al-Ge [20–22]. The favourite valence electron concentration for *hP*1 structure is 3.7 – 4 electrons per atom.

In conclusion, the stability of the high pressure structures for phosphorus is attributed to the lowering of the electronic band structure energy due to Brillouin zone – Fermi surface interaction. Appearance of incommensuratly modulated P-IV structure demonstrates enhanced effects of FS-BZ interactions in comparison with the electrostatic energy preferred the high symmetry atomic arrangements . Similar approach can also be applied to the IM structure of the group VI elements.

**Acknowledgments:** The authors gratefully acknowledge Dr. Olga Degtyareva for valuable discussion and comments. This work is supported by the Program "The Matter under High Pressure" of the Russian Academy of Sciences.